\documentclass[prd,aps,preprintnumbers,floats,floatfix,superscriptaddress,preprintnumbers,
showpacs,eqsecnum,nofootinbib,notitlepage,
twocolumn
]{revtex4-1}
\usepackage[utf8]{inputenc}
\usepackage{latexsym,array,theorem,mathrsfs,bm,float}
\usepackage{psfrag}
\usepackage{amsfonts,amsmath,amssymb,latexsym,array,afterpage,
theorem,mathrsfs,bm,float,epsfig,color,graphicx,tabularx,here,multirow}

\newcommand{\nn}{\nonumber \\}

\newcommand{\nablaA}{\overset{\scriptscriptstyle  \Gamma}{\nabla }}

\newcommand{\RA}{\overset{\scriptscriptstyle  \Gamma}{R }}
\newcommand{\GA}{\overset{\scriptscriptstyle  \Gamma}{G }}

\begin{document}
\baselineskip=12pt

%<<<<<<<<<<<<< PREPRINT NUMBER >>>>>>>>>>>>>>>%
%%
\preprint{WU-AP/1805/18}

%<<<<<<<<<<<<< TITLE >>>>>>>>>>>>>>>%
%%
\title{Galileon and generalized Galileon 
with projective invariance\\
in a metric-affine formalism
}
%%
%<<<<<<<<<<<<< AUTHOR >>>>>>>>>>>>>>>%
%%
\author{Katsuki \sc{Aoki}}
\email{katsuki-a12@gravity.phys.waseda.ac.jp}
\author{Keigo \sc{Shimada}}
\email{kshimada@gravity.phys.waseda.ac.jp}
\affiliation{
Department of Physics, Waseda University,
Shinjuku, Tokyo 169-8555, Japan
}

%<<<<<<<<<<<<< DATE >>>>>>>>>>>>>>>%
\date{\today}

%======================================%
%<<<<<<<<<<<<< ABSTRACT >>>>>>>>>>>>>>>%
%======================================%
\begin{abstract}
We study scalar-tensor theories respecting the projective invariance in the metric-affine formalism. The metric-affine formalism is a formulation of gravitational theories such that the metric and the connection are independent variables in the first place. In this formalism, the Einstein-Hilbert action has an additional invariance, called the projective invariance, under a shift of the connection. Respecting this invariance for the construction of the scalar-tensor theories, we find that the Galileon terms in curved spacetime are uniquely specified at least up to quartic order which does not coincide with either the covariant Galileon or the covariantized Galileon. We also find an action in the metric-affine formalism which is equivalent to class ${}^2$N-I/Ia of the quadratic degenerated higher order scalar-tensor (DHOST) theory. The structure of DHOST would become clear in the metric-affine formalism since the equivalent action is just linear in the generalized Galileon terms and non-minimal couplings to the Ricci scalar and the Einstein tensor with independent coefficients. The fine-tuned structure of DHOST is obtained by integrating out the connection.  In these theories, non-minimal couplings between fermionic fields and the scalar field may be predicted. We discuss possible extensions which could involve theories beyond DHOST.
\end{abstract}

%<<<<<<<<<<<<< PACS NUMBER >>>>>>>>>>>>>>>%
%\pacs{04.60.Cf, 04.50.Gh, 04.50.-h, 11.25.-w }

% 04.40.Dg	Relativistic stars: structure, stability, and oscillations (see also 97.60.-s Late stages of stellar evolution)
% 04.50.Gh : higher-dimensional Black holes
% 04.50.-h : Higher-dimensional gravity and other theories of gravity
% 04.60.Cf : gravitational aspects of String theory
% 11.25.-w : Strings and branes
% 04.50.Kd : Modified theories of gravity
% 98.80.-k : Cosmology
% 95.36.+x : Dark energy
%\pacs{}

\maketitle

\section{Introduction}
Einstein's general relativity (GR) is now accepted as the standard theory of gravity which provided an important insight on physics: a gravitational field corresponds to a deviation of the spacetime geometry from the flat spacetime geometry.  Although it was believed that only the (pseudo-)Euclidean geometry is relevant to physics, the idea of GR tells us that the Euclidean geometry is just a special case in physical systems. GR is usually formulated in the (pseudo-)Riemannian geometry in which all intrinsic structure of the geometry is uniquely determined by the metric. The Einstein equation is regarded as the equation of motion of the metric. However, we should emphasize that the Riemannian geometry is still a special case and there is a more general framework of the geometry called the metric-affine geometry (see \cite{Kleyn:2004yj,Blagojevic:2013xpa} for reviews). The structure of the metric-affine geometry is defined in terms of two independent geometrical objects, the metric and the connection, i.e., the quantities defining the inner product and the parallel transport, respectively. Only if one assumes the metric compatibility condition and the torsionless condition (detailed in Sec.~\ref{sec_metric_affine}), the metric-affine geometry is reduced to the Riemannian geometry. A point is that GR formulated in the metric-affine geometry is effectively equivalent to GR in the Riemannian geometry in a vacuum because the metric compatibility condition and the torsionless condition are obtained from the equation of motion of the independent connection~\cite{giachetta1997projective,Sotiriou:2009xt,Dadhich:2010xa}. It is important to stress that, we do not need to assume the Riemannian geometry in the first place to obtain GR.

When one recede from GR, however, the equivalence of the theories in the Riemannian geometry and in the metric-affine geometry cease to exist.  A popular example of this is metric-affine $f(R)$ theories, also sometimes referred to as Palatini $f(R)$ theories~\cite{Sotiriou:2009xt,Sotiriou:2006qn} in which the resulting geometry is either Riemann-Cartan geometry or integrable Weyl geometry when solving the equation of the connection. Such theories differ from their metric formalism counterpart and have been applied in cosmological scenarios (for a review see~\cite{DeFelice:2010aj, Olmo:2011uz}). Further extending the $f(R)$ theories, one could consider metric-affine formalism in $f(R_{\mu\nu})$ theories~\cite{Borowiec:1996kg}, consider gravity coupling with the energy-momentum tensor in $f(R,T)$ theories~\cite{Barrientos:2018cnx} or introduce two curvature tensor, one from the metric and the other from the connection, in hybrid metric-Palatini gravity~\cite{Harko:2011nh}. 

%However one may note that there has been no unified formalism extending beyond the curvature tensors in alternative theories of gravity.

Another way to simply extend GR is introducing a scalar degree of freedom that describes the gravitational field in addition to the tensor degrees of freedom. This is commonly called scalar-tensor theories. Although many of these theories have been proposed, there are unified descriptions of the scalar-tensor theories. The Horndeski theory~\cite{Horndeski:1974wa,Charmousis:2011bf,Deffayet:2011gz,Kobayashi:2011nu,Ezquiaga:2016nqo} is the most general scalar-tensor theory with the equation of motion with at most second derivatives. The assumption on the number of derivatives is imposed to avoid the Ostrogradsky ghost. However, the discovery of the Gleyzes-Langlois-Piazza-Vernizzi (GLPV) theory~\cite{Gleyzes:2014dya,Gleyzes:2014qga} revealed that the assumption of the derivatives is too strong to obtain a general description of the Ostrogradsky ghost-free theories. The currently known most general theory with one scalar is called the degenerated higher order scalar-tensor (DHOST) theory~\cite{Langlois:2015cwa,BenAchour:2016fzp} (see also \cite{Zumalacarregui:2013pma}). Note that these theories are formulated based on the Riemannian geometry. A little attention has been paid to scalar-tensor theories in the metric-affine geometry.

Since the Riemannian geometry is a quite strong assumption for the description of gravitational theories, it would be natural to ask whether a scalar-tensor theory can be reformulated in the metric-affine geometry as with GR and whether there is a theory beyond DHOST or not. We shall call gravitational theories formulated in the Riemannian geometry the theories in the metric formalism and those in the metric-affine geometry the theories in the metric-affine formalism, respectively. In the metric-affine formalism, the Einstein-Hilbert action and the standard matter action have an additional gauge invariance, the projective invariance, under a shift of the connection (given by \eqref{gauge_inv} later). One may discuss both projective invariant scalar-tensor theories and non-projective invariant theories; in the latter case, a constraint on the connection is imposed to eliminate the projective mode~\cite{Sotiriou:2006qn,Li:2012cc}. Hence, we shall focus on the projective invariant case in order not to impose any assumption on the connection. If the Lagrangian does not contain either higher order derivatives of the scalar field or non-minimal couplings to the curvature, the Lagrangian has no additional connection dependence and then it is trivially projective invariant. However, the higher derivatives or the non-minimal couplings yield explicit dependence of the connection and it has not been known how to construct the general projective invariant Lagrangian with such scalar degree of freedom (see \cite{Afonso:2017bxr} for the case of non-minimal couplings to the symmetric part of the Ricci tensor).

%Nevertheless, it is an important symmetry to consider when constructing a metric-affine theory.

%\textcolor{red}{It has been investigated recently that projective invariance allows some theories to gauge out torsion while others do not, which answered the consistency of substituting torsionless conditions apriori~\cite{Afonso:2017bxr}. }

In the present paper, we thus discuss scalar-tensor theories with the projective invariance. The projective invariance leads to a restriction on the form of the higher derivative terms of the scalar field Lagrangian. Indeed, we find that the covariant Galileon terms are uniquely determined by the projective invariance in the metric-affine formalism at least up to the quartic order although those in the metric formalism are not unique. Then, a question arises: can the projective invariance prohibit the appearance of the Ostrogradsky ghost? Since the higher derivative terms in the DHOST theory are fine-tuned to eliminate the ghost, it should be interesting to seek a hidden symmetry to protect the structure of the DHOST theory. We find a projective invariant action in the metric-affine formalism which is equivalent to class ${}^2$N-I/Ia of DHOST in the metric formalism, where ${}^2$N-1 is named by~\cite{Crisostomi:2016czh,BenAchour:2016fzp} and Ia is by~\cite{Achour:2016rkg}, when we use the equation of motion of the connection. The equivalent action is just linear in the Galileon terms and the non-minimal couplings to the Ricci scalar and the Einstein tensor with independent coefficients. However, we also find other projective invariant terms yielding the Ostrogradsky ghost. Therefore, the DHOST theory can be reformulated to be projective invariant but this symmetry cannot prohibit the appearance of the ghost.

The paper is organized as follows. In Sec.~\ref{sec_metric_affine}, we review the basic concepts of metric-affine gravity and how the Riemannian geometry emerges from the metric-affine geometry when casting special conditions to the connection. We will then introduce a symmetry of the connection called projective symmetry which appears when considered the metric-affine formalism of GR. In Sec.~\ref{sec_galileon}, we will formulate Galileon in metric-affine formalism and find that, when projective invariance is assumed the Galileon terms are uniquely determined.  The equation of the connection can be explicitly solved and then by integrating it out, we find an effective description of metric-affine Galileon in Riemannian geometry. In Sec.~\ref{sec_GG}, we go further into considering Generalized Galileons in terms of the metric-affine formalism and find that in an effective description of Riemannian geometry the theory becomes a class ${}^2$N-I/Ia DHOST theory, i.e. the theory has no Ostrogradsky ghost.  Then in Sec.~\ref{sec_higher_orders}, we argue the Lagrangian cubic in the second derivative of the scalar field. Finally, we make summary remarks in the last Sec.~\ref{sec_summary}. In Appendix \ref{appendix}, we discuss generic projective invariant scalar-tensor theories with at most quadratic in the connection and show the ghost-free conditions and the classifications of generic theories.

\section{Metric-affine formalism}
\label{sec_metric_affine}
\subsection{Metric-affine, Riemann-Cartan, Riemannian, and Euclidean geometries}
The intrinsic structure of the metric-affine geometry is defined in terms of the metric $g_{\mu\nu}$ and the connection $\Gamma^{\mu}_{\alpha\beta}$. We should emphasize that the connection and the metric are independent geometrical objects in the first place. For mathematical rigorousness of this geometry see for example ~\cite{Kleyn:2004yj}. The covariant derivatives for a vector are defined by
\begin{align}
\nablaA_{\alpha}A^{\mu}
&=\partial_{\alpha}A^{\mu}+\Gamma^{\mu}_{\beta\alpha}A^{\beta}
\,,
\\
\nablaA{}_{\alpha}A_{\mu}&=\partial_{\alpha}A_{\mu}-\Gamma^{\beta}_{\mu\alpha}A_{\beta}\,.
\end{align}
In a manifold with a metric and a connection, there are three tensors that characterize the geometry: Riemann curvature, torsion, and non-metricity. These are defined by
\begin{align}
\RA{}^{ \mu}{}_{\nu\alpha\beta}(\Gamma)&:=\partial_{\alpha}\Gamma^{\mu}_{\nu\beta}-\partial_{\beta}\Gamma^{\mu}_{\nu\alpha}
+\Gamma^{\mu}_{\sigma\alpha } \Gamma^{\sigma}_{\nu\beta }-\Gamma^{\mu}_{ \sigma \beta} \Gamma^{\sigma}_{\nu\alpha } \,,
\nn
T^{\mu}{}_{\alpha\beta}&:=\Gamma^{\mu}_{\beta\alpha}-\Gamma^{\mu}_{\alpha\beta}
\,,\\
Q_{\mu}{}^{\alpha\beta}&:=\nablaA{}_{\mu}g^{\alpha\beta}\,.
\end{align}

In four dimensions, the metric has $10$ independent components and the connection has $64$ independent components, respectively. To simplify the structure of the geometry, one may assume the metric compatibility condition
\begin{align}
Q_{\mu}{}^{\alpha\beta}=0\,.
\end{align}
This condition is obtained if we demand that the inner product of two vectors is preserved under the parallel transport because
\begin{align}
&g_{\mu\nu}(x+dx)A^{\mu}_{\rm PT}(x+dx)B^{\nu}_{\rm PT}(x+dx)
\nn
&-g_{\mu\nu}(x)A^{\mu}(x)B^{\nu}(x)
\nn
=&-Q_{\alpha\mu\nu}A^{\mu}B^{\nu}dx^{\alpha}\,,
\end{align}
where
\begin{align}
g_{\mu\nu}(x+dx)&=g_{\mu\nu}(x)+\partial_{\alpha}g_{\mu\nu}(x)dx^{\alpha}\,,
\\
A^{\mu}_{\rm PT}(x+dx)&=A^{\mu}(x)-\Gamma^{\mu}_{\alpha\beta}(x)A^{\alpha}(x)dx^{\beta}\,.
\end{align}
When the covariant derivative is metric compatible, the connection is called a metric connection.
Then, the geometry is reduced to the Riemann-Cartan geometry in which the connection is given by
\begin{align}
\Gamma^{\mu}_{\alpha\beta}=\left\{ {}^{\,\, \mu}_{\alpha\beta} \right\} -\frac{1}{2}\left(T^{\mu}{}_{\alpha\beta}-T_{\beta}{}^{\mu}{}_{\alpha}+T_{\alpha\beta}{}^{\mu} \right) \,, 
\end{align}
where $\left\{ {}^{\,\, \mu}_{\alpha\beta} \right\}$ is the Levi-Civita connection defined by
\begin{align}
\left\{ {}^{\,\, \mu}_{\alpha\beta} \right\} := \frac{1}{2}g^{\mu\nu}(\partial_{\alpha} g_{\beta\nu}+\partial_{\beta}g_{\alpha\nu}-\partial_{\nu}g_{\alpha\beta} ) \,.
\end{align}

Furthermore, one may assume the torsionless condition
\begin{align}
T^{\mu}{}_{\alpha\beta}=0
\,,
\end{align}
and now parallel displacement is fully characterized by the Riemann tensor.  As a result, we obtain the Riemannian geometry in which the connection is uniquely determined to be the Levi-Civita connection. The 64 independent components are now fixed and then the structure of the geometry is determined by the metric only.

When we further assume
\begin{align}
\RA{}^{\mu}{}_{\nu\alpha\beta}=0\,,
\end{align}
the Euclidean geometry is obtained.

\subsection{Metric-affine formalism of GR}
When a gravitational theory is formulated in the Riemannian geometry, the independent variable is the metric only. This formalism of these gravitational theories is called the metric formalism. However, as discussed above, the general geometry does not require the connection is given by the Levi-Civita connection. Hence, it would be natural to promote that the metric and the connection are independent variables in the first place and a gravitational theory dynamically determines not only the metric but also the connection. This is called the metric-affine formalism.

It is known that the Einstein-Hilbert (EH) action in the metric-affine formalism is equivalent to that in the metric formalism in vacuum~\cite{giachetta1997projective,Sotiriou:2009xt,Dadhich:2010xa}. Let us consider the EH action
\begin{align}
S_{\rm EH}(g,\Gamma)&=\int d^4x  \sqrt{-g} \mathcal{L}_{\rm EH}
\,, 
\nn
\mathcal{L}_{\rm EH}(g,\Gamma)&= \frac{M_{\rm pl}^2}{2}g^{\mu\nu} \RA{}_{\mu\nu}
\end{align}
where $\RA{}_{\mu\nu}=\RA{}^{\alpha}{}_{\mu\alpha\nu}$. Since the metric and the connection are independent variables in the metric-affine formalism, the variation of the EH action leads to two independent equations. To take the variation, we have to take care of the fact that the connection is not a tensor. The easiest way is to express the connection as
\begin{align}
\Gamma^{\mu}_{\alpha\beta}=\left\{ {}^{\,\, \mu}_{\alpha\beta} \right\} + \kappa^{\mu}{}_{\alpha\beta}
\,,
\end{align}
and to regard the distortion tensor $\kappa^{\mu}{}_{\alpha\beta}$ as the independent variable instead of the connection itself. Then, the EH action is rewritten by
\begin{align}
\mathcal{L}_{\rm EH}(g,\Gamma)=\frac{M_{\rm pl}^2}{2} \left( R(g) 
+ \kappa^{\alpha}{}_{\beta\alpha}\kappa^{\beta\gamma}{}_{\gamma} - \kappa^{\alpha\beta\gamma} \kappa_{\beta\gamma\alpha}
\right)\,, \label{EH_kappa}
\end{align}
where $R(g)$ is the Ricci scalar constructed by the Levi-Civita connection.

Before proceeding with the variation, we note that the EH action has an additional gauge invariance, called the projective invariance, under the transformation
\begin{align}
\Gamma^{\mu}_{\alpha\beta}\rightarrow \Gamma^{\mu}_{\alpha\beta}+\delta^{\mu}_{\alpha}U_{\beta}\,, \label{gauge_inv}
\end{align}
for an arbitrary vector $U_{\alpha}(x)$. Geometrically, the projective transformation is a change of the connection which preserves the geodesic equation
\begin{align}
\frac{d^2 x^{\mu}}{d \lambda^2}+\Gamma^{\mu}_{\alpha\beta} \frac{dx^{\alpha}}{d\lambda} \frac{dx^{\beta}}{d\lambda}=0
\end{align}
up to the redefinition of the affine parameter $\lambda\to \tilde \lambda(\lambda)$~\cite{riccischouten,veblen1926projective}. Note that the general projective transformation is given by
\begin{align}
\Gamma^{\mu}_{\alpha\beta}\rightarrow \Gamma^{\mu}_{\alpha\beta}+\delta^{\mu}_{\alpha}U_{\beta} +\delta^{\mu}_{\beta}V_{\alpha}\,,
\end{align}
with two arbitrary vectors $U_{\alpha}$ and $V_{\beta}$. The transformation \eqref{gauge_inv} is a special class of the projective transformation which also preserves the angle between two vectors under the parallel transport since the non-metricity tensor is transformed as
\begin{align}
Q_{\mu}{}^{\alpha\beta} \rightarrow Q_{\mu}{}^{\alpha\beta}+2U_{\mu}g^{\alpha\beta} \,.
\end{align}
In the present paper, we just call the transformation \eqref{gauge_inv} the projective transformation and the invariance under it the projective invariance, respectively. For further explanation of geometrical characteristics of projective transformation see section VI of the textbook ~\cite{riccischouten}.

After introducing the distortion tensor $\kappa^{\mu}{}_{\alpha\beta}$, the projective invariance is cast in the invariance under 
\begin{align} 
\kappa^{\mu}{}_{\alpha\beta}\rightarrow \kappa^{\mu}{}_{\alpha\beta}+\delta^{\mu}_{\alpha}U_{\beta}\,, \label{tras_kappa}
\end{align}
which results the identity
\begin{align}
\delta^\mu_\alpha \frac{\delta S_{\rm EH}}{\delta {\kappa^\mu}_{\alpha\beta}} \equiv 0 \,.
\end{align}

Since the distortion tensor is a non-dynamical field in the action \eqref{EH_kappa}, the distortion tensor can be integrated out. The variation with respect to  $\kappa^{\mu}{}_{\alpha\beta}$ yields the solution,
\begin{align}
\kappa^{\mu}{}_{\alpha\beta} = 0\,,
\end{align}
up to the gauge freedom. Note that, although the solution of $\kappa^{\mu}{}_{\alpha\beta}$ is not uniquely determined due to the freedom of the projective transformation, we just omit the gauge mode because the gauge mode does not affect the motion of the physical variables. As a result, the EH action in the metric-affine formalism coincides with the EH action in the metric formalism after integrating out the distortion tensor.

\subsection{Coupling to matter}

The equivalence between the metric-affine formalism and the metric formalism must not be true in general if we add either higher curvature terms or a matter field. In the present paper, we shall consider a sufficiently low energy scale so that higher curvature terms can be ignored, and then only focus on the latter one, the inclusion of matter, which enables us to integrate out the distortion tensor since $\kappa$ is still a non-dynamical field.

In this section, we discuss a minimal scalar field $\phi$, a vector field $A^{\mu}$, and a Dirac field $\psi$. We consider the action
\begin{align}
S=S_{\rm EH}(g,\Gamma)+S_{\rm m}(g,\Gamma,\phi,A,\psi)\,,
\end{align}
where $S_{\rm m}$ is a matter action which generally contains the connection as well as the metric. Similarly to the previous case, we can introduce the distortion tensor, 
\begin{align}
S=S_{\rm EH}(g,\kappa)+S_{\rm m}(g,\kappa,\phi,A,\psi)\,.
\end{align}
The matter fields can be a source of the distortion tensor and then $\kappa^{\mu}{}_{\alpha\beta}=0$ up to the gauge mode is no longer the solution to the equation of motion, in general.

We note that the projective invariance still holds even if we add the standard matter fields.
The minimal kinetic term of the scalar field is given by $-\frac{1}{2}g^{\mu\nu}\partial_{\mu} \phi \partial_{\nu} \phi$ which is manifestly projective invariant. As for the vectors, since the appropriate definition of the covariant field strength of the vector field is
\begin{align}
F_{\mu\nu}:=\partial_{\mu}A_{\nu}-\partial_{\nu}A_{\mu}\,, \label{fmunu}
\end{align}
the vector field does not couple with the distortion tensor and then the action of the vector field is invariant under \eqref{tras_kappa} which is also true for the Yang-Mills fields. One could propose that the covariant field strength is actually written with the covariant derivative of the connection as,
\begin{align}
{\overset{\Gamma}F}_{\mu\nu}:=\nablaA{}_{\mu}A_\nu-\nablaA{}_{\nu}A_\mu.
\end{align}
However, first this field strength is not $U(1)$ invariant, and second, when considered in the language of differential forms the field strength is $F=dA$ and this in tensor form is \eqref{fmunu}. For more discussion see for example~\cite{Hehl:1976kj,Hojman:1978yz}.

While the bosonic field do not couple to the distortion tensor, the Dirac field has a coupling to $\kappa$ and thus the Dirac field can be a source of $\kappa$. To discuss the Dirac field, we have to introduce the tetrad $e^a_{\mu}$ and the spin connection $\omega^{ab}{}_{\mu}$. We assume the tetrad postulate which reads that the connection $\Gamma$ and the spin connection $\omega$ represent the same geometrical object, i.e.,
\begin{align}
A^a_{\rm PT}(x+dx)=e^a_{\mu}(x+dx) A^{\mu}_{\rm PT}(x+dx)\,,
\end{align}
where 
\begin{align}
A^a_{\rm PT}(x+dx)&=A^a(x)-\omega^a{}_{b\mu}A^b(x)dx^{\mu}
\,, \\
e^a_{\mu}(x+dx)&=e^a_{\mu}+\partial_{\alpha}e^a_{\mu}(x)dx^{\alpha}\,.
\end{align}
Then, the variables $(g,\Gamma)$ and $(e,\omega)$ are related by
\begin{align}
g_{\mu\nu}&=\eta_{ab}e^a_{\mu}e^b_{\nu}\,, \\
\nablaA{}_{\mu}e^a_{\nu}&=\partial_{\mu}e^a_{\nu}-\Gamma^{\alpha}_{\nu\mu}e^a_{\alpha}+\omega^{a}{}_{b \mu}e^b_{\nu}=0
\,.
\end{align}
The second equation leads to that the spin connection can be written by
\begin{align}
\omega^{ab}{}_{\mu}=\Delta^{ab}{}_{\mu}+\kappa^{ab}{}_{\mu}
\end{align}
where $\Delta^{ab}{}_{\mu}$ are the Ricci rotation coefficients and
\begin{align}
\kappa^{ab}{}_{\mu}=e^a_{\alpha}e^b_{\beta}\kappa^{\alpha\beta}{}_{\mu} \,.
\end{align}
The covariant derivative of the Dirac field $\psi$ is then given by
\begin{align}
\nablaA{}_{\mu} \psi = \left( \partial_{\mu} + \frac{1}{8}\omega^{ab}{}_{\mu} [\gamma_{a}, \gamma_b ] \right) \psi
 \,, \label{D_psi}
\end{align}
where $\gamma_a$ is the gamma matrix with $\{ \gamma_a ,\gamma_b \}=-2\eta_{ab}$. Since $[\gamma_{a}, \gamma_b ]$ is antisymmetric for the indices $a,b$, \eqref{D_psi} is projective invariant.

The Dirac field Lagrangian in the metric-affine geometry may be
\begin{align}
\mathcal{L}_D = \frac{i}{2} \left( \bar{\psi} \gamma^{\mu} \nablaA{}_{\mu} \psi - (\nablaA{}_{\mu} \bar{\psi}) \gamma^{\mu} \psi 
\right)-m\bar{\psi}\psi\,, \label{Dirac1}
\end{align}
or
\begin{align}
\mathcal{L}_D' =i \bar{\psi} \gamma^{\mu} \nablaA{}_{\mu} \psi -m\bar{\psi}\psi\,, \label{Dirac2}
\end{align}
where $\gamma^{\mu}=e^{\mu}_a \gamma^a$. Admitting the equivalence upon the integration by parts, only difference between \eqref{Dirac1} and \eqref{Dirac2} is the coupling to the distorsion tensor.
%In the Riemannian geometry, two Lagrangian are equivalent up to the freedom of the total divergence terms, while two are not in the metric-affine geometry. Indeed, 
The interaction terms are given by
\begin{align}
\mathcal{L}_{\rm int}=-\frac{1}{4}\epsilon^{\alpha\beta\gamma\delta}\kappa_{\alpha\beta\gamma}j^5_{\delta} \,,
\end{align}
and
\begin{align}
\mathcal{L}_{\rm int}'=-\frac{1}{4}\epsilon^{\alpha\beta\gamma\delta}\kappa_{\alpha\beta\gamma}j^5_{\delta}
+ \frac{i}{2} \kappa^{[\alpha\beta]}{}_{\beta} j_{\alpha} \,, \label{int}
\end{align}
respectively, where
\begin{align}
j_{\mu}=\bar{\psi}\gamma_{\mu}\psi\,, \quad j_5^{\mu}=\bar{\psi} \gamma_{\mu} \gamma^5\psi \,,
\end{align}
with $\gamma^5=-\frac{i}{4!}\epsilon^{\alpha\beta\gamma\delta}\gamma_{\alpha}\gamma_{\beta}\gamma_{\gamma}\gamma_{\delta}$ and $\epsilon^{\alpha\beta\gamma\delta}$ is the Levi-Civita tensor.
Due to the coupling to the distortion tensor, $\kappa^{\mu}{}_{\alpha\beta}=0$ is not a solution if the Dirac field exists for both cases. Then, the equivalence between the metric one and the metric-affine one does not hold which is well-known in the context of the Einstein-Cartan-Sciama-Kibble theory.

\section{Galileon in metric-affine formalism}
\label{sec_galileon}
In the previous section, we discussed that the standard Lagrangian including matter as well as gravity is projective invariant. Hence, it would be natural to ask whether a non-standard Lagrangian can be projective invariant or not. We consider a scalar field and assume the projective invariance for the construction of the scalar field Lagrangian.

In this section, we study the Galileon scalar field. In the flat spacetime, the action of the Galileon scalar field is specified to enjoy the Galileon invariance
\begin{align}
\phi \rightarrow \phi + b_{\mu}x^{\mu}+c
\end{align}
where $b_{\mu}$ and $c$ are constant parameters~\cite{Nicolis:2008in}. The flat spacetime Lagrangian of the Galileon scalar is given by
\begin{align}
\mathcal{L}=\sum_{n\geq 2}^5 \frac{c_n}{\Lambda_3^{3(n-2)}} \mathcal{L}_n^{\rm gal}
\end{align}
with
\begin{align}
\mathcal{L}_2^{\rm gal} &:= \epsilon^{\alpha\beta\gamma\delta} \epsilon^{\alpha'}{}_{ \beta\gamma\delta} \partial_{\alpha} \phi \partial_{\alpha'} \phi \label{gal2}
\,, \\
\mathcal{L}_3^{\rm gal} &:= \epsilon^{\alpha\beta\gamma\delta} \epsilon^{\alpha'\beta' }{}_{\gamma\delta} \partial_{\alpha} \phi \partial_{\alpha'} \phi \partial_{\beta}\partial_{\beta'}\phi 
\,, \\
\mathcal{L}_4^{\rm gal} &:= \epsilon^{\alpha\beta\gamma\delta} \epsilon^{\alpha'\beta'\gamma'}{}_{\delta} \partial_{\alpha} \phi \partial_{\alpha'} \phi \partial_{\beta}\partial_{\beta'}\phi \partial_{\gamma}\partial_{\gamma'} \phi
\,, \\
\mathcal{L}_5^{\rm gal} &:= \epsilon^{\alpha\beta\gamma\delta} \epsilon^{\alpha'\beta'\gamma'\delta'} \partial_{\alpha} \phi \partial_{\alpha'} \phi \partial_{\beta}\partial_{\beta'}\phi \partial_{\gamma}\partial_{\gamma'} \phi \partial_{\delta} \partial_{\delta'} \phi
\,, \label{gal5}
\end{align}
where $c_n$ are dimensionless constants and $\Lambda_3$ represents the strong coupling scale. We note that the Galileon terms can be expressed by
\begin{align}
\mathcal{L}_2^{\rm gal} &= -6(\partial \phi)^2 \,, \\
\mathcal{L}_3^{\rm gal} &= \frac{1}{2}\partial_{\mu}\phi \partial^{\mu} \phi \epsilon^{\alpha\beta\gamma\delta} \epsilon^{\alpha'}{}_{ \beta\gamma\delta} \partial_{\alpha' }\partial_{\alpha} \phi \,, \nn
&= - (\partial \phi)^2  \Box \phi \,,
\\
\mathcal{L}_4^{\rm gal} &= \partial_{\mu}\phi \partial^{\mu} \phi \epsilon^{\alpha\beta\gamma\delta} \epsilon^{\alpha' \beta' }{}_{\gamma\delta} \partial_{\alpha' }\partial_{\alpha} \phi \partial_{\beta'}\partial_{\beta} \phi  \nn
&=-2 (\partial \phi)^2 \left[ (\Box \phi)^2-( \partial_{\alpha }\partial_{\beta} \phi)^2 \right]
\,,
\\
\mathcal{L}_5^{\rm gal} &= \frac{5}{2}\partial_{\mu}\phi \partial^{\mu} \phi \epsilon^{\alpha\beta\gamma\delta} \epsilon^{\alpha' \beta' \gamma' }{}_{\delta} \partial_{\alpha' }\partial_{\alpha} \phi \partial_{\beta'}\partial_{\beta} \phi \partial_{\gamma'}\partial_{\gamma} \phi
\nn
&= -\frac{5}{2} (\partial \phi)^2 \left[ (\Box \phi)^3-3\Box \phi (\partial_{\alpha}\partial_{\beta}\phi)^2 +2 (\partial_{\alpha}\partial_{\beta}\phi)^3 \right]\,,
\end{align}
after taking the integration by parts. As a result, the Galileon terms in the flat spacetime are schematically given by
\begin{align}
\mathcal{L}^{\rm gal}_n &=\epsilon \epsilon (\partial \phi)^2 (\partial \partial \phi)^{n-2}
\nn
&=(\partial \phi)^2 \epsilon \epsilon (\partial \partial \phi)^{n-2} +{\rm ~total~divergence}\,.
\end{align}

We then consider the Galileon field in the curved spacetime. In the metric formalism, there are two ways to covariantize the Galileon field: the covariant Galileon~\cite{Deffayet:2009wt} and the covariantized Galileon~\cite{Nicolis:2008in}, respectively. The covariant Galileon is based on the form $(\partial \phi)^2 \epsilon \epsilon (\partial \partial \phi)^{n-2}$, while the covariantized Galileon is given by the form $\epsilon \epsilon (\partial \phi)^2 (\partial \partial \phi)^{n-2}$. The equivalence between them no longer holds in the curved spacetime and then the covariant Galileon and the covariantized Galileon are different theories. Indeed, the covariant Galileon is a theory in the class of Horndeski theory and the covariantized Galileon is in the class of GLPV theory, respectively. 

When we consider the metric-affine formalism instead of the metric formalism, the covariant theory must be invariant under the projective transformation\footnote{Galileon terms in the metric-affine formalism were considered in~\cite{Li:2012cc} without respecting the projective invariance. They instead introduce an additional constraint on the connection to eliminate the projective mode.}. 
Here, we assume the Galileon terms are purely constructed by the covariant derivatives of the scalar field and the metric, i.e., 
\begin{align}
\mathcal{L}_n^{\rm gal\Gamma}=\mathcal{L}_n^{\rm gal\Gamma}(g,\phi,\nablaA_{\mu}\phi,\nablaA_{\mu}\nablaA{}_{\nu}\phi)\,.
\label{minimal_gal}
\end{align}
 This assumption is a minimal way to introduce derivative interactions of a scalar field in the metric-affine formalism. Since
\begin{align}
\nablaA{}^{\mu} \nablaA{}^{\nu}\phi &=g^{\mu\alpha}g^{\nu\beta} \nablaA{}_{\alpha}\nablaA{}_{\beta}\phi+Q^{\mu\nu\gamma}\nablaA{}_{\gamma}\phi 
\nn
&\neq g^{\mu\alpha}g^{\nu\beta} \nablaA{}_{\alpha}\nablaA{}_{\beta}\phi \,,
\end{align}
due to the non-metricity tensor where $\nablaA{}^{\mu}:=g^{\mu\nu}\nablaA{}_{\nu}$, a Lagrangian containing $\nablaA{}^{\mu} \nablaA{}^{\nu}\phi $ implicitly contains the non-metricity tensor in addition to the covariant derivatives of the scalar field, i.e., 
\begin{align}
&\mathcal{L}(g,\phi,\nablaA_{\mu}\phi,\nablaA_{\mu}\nablaA{}_{\nu}\phi,\nablaA{}^{\mu}\nablaA{}^{\nu}\phi)
\nn
&=\mathcal{L}(g,\phi,\nablaA_{\mu}\phi,\nablaA_{\mu}\nablaA{}_{\nu}\phi,Q_{\alpha}{}^{\beta\gamma})\,.
\end{align}
Supposing the minimal derivative interactions of the scalar field \eqref{minimal_gal}, we find that the Lagrangian of the covariant Galileon terms is uniquely specified by the projective invariance at least up to the quartic order.

The form $\epsilon \epsilon (\partial \phi)^2 (\partial \partial \phi)^{n-2}$ is projective invariant, while $(\partial \phi)^2 \epsilon \epsilon (\partial \partial \phi)^{n-2}$ is not. In this way, we obtain all projective invariant terms which reproduce \eqref{gal2}-\eqref{gal5} by replacing $\nablaA{}_{\mu}$ with $\partial_{\mu}$ except the freedom of the integration by parts. The projective invariant Galileon terms are
\begin{align}
\mathcal{L}_2^{\rm gal \Gamma} &= \epsilon^{\alpha\beta\gamma\delta} \epsilon^{\alpha'}{}_{ \beta\gamma\delta} \nablaA{}_{\alpha} \phi \nablaA{}_{\alpha'} \phi
\,, \label{galG2} \\
\mathcal{L}_3^{\rm gal \Gamma} &= \epsilon^{\alpha\beta\gamma\delta} \epsilon^{\alpha'\beta' }{}_{\gamma\delta} \nablaA{}_{\alpha} \phi \nablaA{}_{\alpha'} \phi  \nablaA{}_{\beta} \nablaA{}_{\beta'}\phi 
\,, \\
\mathcal{L}_4^{\rm gal \Gamma} &= \epsilon^{\alpha\beta\gamma\delta} \epsilon^{\alpha'\beta'\gamma'}{}_{\delta} \nablaA{}_{\alpha} \phi \nablaA{}_{\alpha'} \phi \nablaA{}_{\beta} \nablaA{}_{\beta'}\phi \nablaA{}_{\gamma} \nablaA{}_{\gamma'} \phi
\,, \label{galG4} \\
\mathcal{L}_5^{\rm gal \Gamma} &= \epsilon^{\alpha\beta\gamma\delta} \epsilon^{\alpha'\beta'\gamma'\delta'} \nablaA{}_{\alpha} \phi \nablaA{}_{\alpha'} \phi \nablaA{}_{\beta} \nablaA{}_{\beta'}\phi \nablaA{}_{\gamma} \nablaA{}_{\gamma'} \phi \nablaA{}_{\delta} \nablaA{}_{\delta'} \phi 
\,,  \label{galG5} 
\end{align}
and
\begin{align}
\mathcal{L}_4^{\rm gal \Gamma'} &= \epsilon^{\alpha\beta\gamma\delta} \epsilon^{\alpha'\beta'\gamma'}{}_{\delta} \nablaA{}_{\alpha} \phi \nablaA{}_{\alpha'} \phi \nablaA{}_{\beta} \nablaA{}_{\beta'}\phi \nablaA{}_{\gamma'} \nablaA{}_{\gamma} \phi
\,, \label{galG4'}\\
\mathcal{L}_5^{\rm gal \Gamma'} &= \epsilon^{\alpha\beta\gamma\delta} \epsilon^{\alpha'\beta'\gamma'\delta'} \nablaA{}_{\alpha} \phi \nablaA{}_{\alpha'} \phi \nablaA{}_{\beta} \nablaA{}_{\beta'}\phi \nablaA{}_{\gamma} \nablaA{}_{\gamma'} \phi \nablaA{}_{\delta'} \nablaA{}_{\delta} \phi 
\,,  \label{galG5'}
\end{align}
where we note that the second derivative has no symmetric indices, 
\begin{align}
2\nablaA{}_{[\mu}\nablaA{}_{\nu]}\phi=-T^{\alpha}{}_{\mu\nu}\partial_{\alpha}\phi \neq 0\,,
\end{align} 
and then \eqref{galG4'} (and \eqref{galG5'}) potentially differs from \eqref{galG4} (and \eqref{galG5}). However, as shown in Appendix \ref{appendix}, \eqref{galG4} and \eqref{galG4'} give the same result and thus it is sufficient to consider only one of them. We may conjecture that \eqref{galG5} and \eqref{galG5'} lead to the same result as well and then we shall ignore \eqref{galG4'} and \eqref{galG5'}. Adding the EH action, the total action of the covariant Galileon in the metric-affine formalism is then given by
\begin{align}
\mathcal{L}(g,\Gamma,\phi)= \frac{M_{\rm pl}^2}{2} g^{\mu\nu} \RA{}_{\mu\nu}+ \sum_{n\geq 2}^5 \frac{c_n}{\Lambda_3^{3(n-2)}} \mathcal{L}_n^{\rm gal\Gamma}  \,. \label{galA}
\end{align}
At least up to the quartic order, the action \eqref{galA} is the unique covariant Galileon theory in the metric-affine formalism.

Just for simplicity, we consider the Galileon terms up to $n=4$ hereafter. This is because the term $\mathcal{L}_5^{\rm gal\Gamma}$ is cubic in the connection and then the equation of motion of the connection becomes nonlinear. We could not find an explicit solution in the nonlinear case. On the other hand, we can find explicit solutions of the connection up to $n=4$.

As performed in the previous section, we introduce the distortion tensor $\kappa^{\mu}{}_{\alpha\beta}$ and integrate it out to obtain the effective action of the metric formalism. The variation with respect to $\kappa^{\mu}{}_{\alpha\beta}$ yields the solution
\begin{align}
\kappa^{\mu}{}_{\alpha\beta}&=-\frac{1}{M_{\rm pl}^2(1+2c_4X^2/\Lambda_2^8)}
\nn
& \times
\Biggl[\frac{c_3}{\Lambda_3^3}\left( X\delta^{\mu}_{\beta}\phi_{\alpha} -X\phi^{\mu} g_{\alpha\beta}+2\phi^{\mu}\phi_{\alpha}\phi_{\beta }\right)
\nn 
&\qquad
+\frac{2c_4}{\Lambda_3^6}\Big\{ 2X \phi^{\mu}{}_{(\alpha}\phi_{\beta)} -X \phi^{\mu}\phi_{\alpha\beta}
\nn
&\qquad \qquad \quad
+\phi_{\alpha}\phi_{\beta}
(\phi^{\mu}\phi^{\gamma}{}_{\gamma}-2\phi^{\mu\gamma}\phi_{\gamma} ) \Big\} \Biggl]\,, \label{gal_kappa}
\end{align}
up to the gauge freedom where we have introduced the notation $\phi_{\mu}=\nabla_{\mu}\phi,\phi_{\mu\nu}=\nabla_{\mu}\nabla_{\nu}\phi,X=\phi^{\mu}\phi_{\mu}$ and $\nabla_{\mu}$ is the covariant derivative with respect to the Levi-Civita connection. The scale $\Lambda_2$ is defined by
\begin{align}
\Lambda_2^4=\Lambda_3^3 M_{\rm pl} \,.
\end{align}
The result explicitly shows that the distortion is no longer zero due to the Galileon scalar field. Unlike the Metric-affine $f(R)$ theories neither the torsionless condition or metric compatibility could be obtained by a suitable fixing of the projective gauge. Substituting it to \eqref{galA}, the resultant action is given by
\begin{align}
\mathcal{L}
&= \frac{M_{\rm pl}^2}{2}R(g)+ \frac{3(c_3^2-4c_2c_4) X^3 /\Lambda_2^8 }{1+2c_4 X^2/\Lambda_2^8} 
\nn
&+\frac{1}{1+2c_4 X^2 /\Lambda_2^8 } \left(c_2 \mathcal{L}_2^{{\rm gal}g}+\frac{c_3}{\Lambda_3^3} \mathcal{L}_3^{{\rm gal}g}+ \frac{c_4}{\Lambda_3^6} \mathcal{L}_4^{{\rm gal}g} \right) \,. \label{cov_gal}
\end{align}
where 
\begin{align}
\mathcal{L}_2^{{\rm gal }g} &= 
\epsilon^{\alpha\beta\gamma\delta} \epsilon^{\alpha'\beta' }{}_{\gamma\delta} \phi_{\alpha} \phi_{\alpha'}  
\,, \\
\mathcal{L}_3^{{\rm gal }g} &= 
\epsilon^{\alpha\beta\gamma\delta} \epsilon^{\alpha' }{}_{\beta' \gamma\delta} \phi_{\alpha} \phi_{\alpha'}  \phi_{\beta \beta'}  
\,, \\
\mathcal{L}_4^{{\rm gal }g} &= \epsilon^{\alpha\beta\gamma\delta} \epsilon^{\alpha'\beta'\gamma'}{}_{\delta} \phi_{\alpha} \phi_{\alpha'} \phi_{\beta \beta'}\phi_{\gamma\gamma'} 
\,.
\end{align}
Therefore, the covariant Galileon in the metric-affine formalism \eqref{galA} is not equivalent to either the covariant Galileon or the covariantized Galileon in the metric formalism. Although \eqref{cov_gal} could be approximated by the covariantized Galileon in the scales below $\Lambda_2$, the deviation becomes relevant when
\begin{align}
|X| \gtrsim \Lambda_2^4\,.
\end{align}
The new scale $\Lambda_2$ naturally arises in the covariant Galileon in the metric-affine formalism.

We thus have three theories of Galileon: the covariant Galileon and the covariantized Galileon in the metric formalism, and the projective invariant Galileon in the metric-affine formalism (or its equivalent form \eqref{cov_gal} after integrating out $\kappa$). When we ignore gravity, i.e., taking the limit $M_{\rm pl} \rightarrow \infty$, all of them reduce to the flat Galileon; however, they do not coincide with each other when gravity is included. 

Since the distortion tensor $\kappa$ is no longer zero, the Galileon field may non-minimally couple with the Dirac field after the integrating out $\kappa$ while $\phi$ does not directly couple with the vector field. Note that the totally antisymmetric part of $\kappa$ is zero and thus the non-minimal coupling does not exist if the Dirac field Lagrangian is given by \eqref{Dirac1}. On the other hand, as for \eqref{Dirac2}, the interaction terms are given by\footnote{When the Dirac field is introduced, Eq.~\eqref{gal_kappa} is not a solution because the Dirac field contributes to the equation of motion of $\kappa$. Nonetheless, in cosmological situations such that the Galileon dominates the universe while the Dirac field can be treated as a test field, Eq.~\eqref{gal_kappa} may be used.}
\begin{align}
\mathcal{L}'_{\rm int}&= \frac{i}{M_{\rm pl}^2( 1+2c_4X^2/\Lambda^8_2) } j_{\alpha}
\nn
&\times 
\Biggl[
\frac{3c_3}{2\Lambda_3^3}X\phi^{\alpha}
+\frac{c_4}{\Lambda_3^6}
\left( X\phi^{\alpha}\phi^{\beta}_{\beta}-\phi^{\alpha} \phi^{\beta\gamma}\phi_{\beta}\phi_{\gamma} \right) \Biggl]\,.
\end{align}
The Galileon in the metric-affine formalism may predict that the bosons (the minimal scalar and vector fields) and the fermions have different couplings to $\phi$ although depending on the definition of the Dirac field Lagrangian.

\section{Generalized Galileon in metric-affine formalism is DHOST}
\label{sec_GG}
\subsection{Equivalent Lagrangian to class ${}^2$N-I/Ia of DHOST}
In the metric formalism, the known most general framework of the scalar-tensor theories without the Ostrogradsky ghost is the DHOST theory. In the DHOST theory, we require a fine-tuning between the coefficients in front of the non-minimal coupling to the curvature and the higher derivatives of the scalar field in order to eliminate the Ostrogradsky ghost. The conditions are called the degeneracy conditions~\cite{Langlois:2015cwa}.

In the metric-affine formalism, we find that the equivalent Lagrangian to (class ${}^2$N-I/Ia of) the DHOST theory is given by
\begin{align}
\mathcal{L}(g,\Gamma,\phi)
&=
f_1(\phi,X) g^{\mu\nu} \RA{}_{\mu\nu} + f_2(\phi,X) \GA{}^{\mu\nu} \nablaA{}_{\mu} \phi \nablaA{}_{\nu} \phi 
\nn
& +F_2(\phi,X) +F_3(\phi,X)\mathcal{L}_3^{\rm gal\Gamma}
+F_4(\phi,X) \mathcal{L}_4^{\rm gal\Gamma}
, \label{generalized_g}
\end{align}
where $f_1,f_2,F_2,F_3,F_4$ are arbitrary functions of $\phi$ and $X:=g^{\mu\nu}\partial_{\mu}\phi \partial_{\nu} \phi$. The last three terms are generalization of the Galileon terms. The first two terms are the non-minimal couplings to the Ricci scalar and the Einstein tensor, respectively, where the Einstein tensor is now defined by
\begin{align}
\GA{}^{\alpha\beta}:=\frac{1}{4} \epsilon^{\gamma \alpha \mu\nu }\epsilon_{\gamma}{}^{\beta\mu'\nu'}\RA{}_{\mu\nu\mu'\nu'} \,.
\end{align}
The action \eqref{generalized_g} can be thus regarded as the straightforward generalization of the Galileon field in the metric-affine formalism including the non-minimal couplings to the curvature.

After integrating out the distortion tensor, we obtain
\begin{align}
\mathcal{L}& =f R(g)+P+Q_1 g^{\mu\nu} \phi_{\mu\nu} +Q_2\phi^{\mu} \phi_{\mu\nu}\phi^{\nu} 
\nn &\quad
+ C^{\mu\nu,\rho\sigma}  \phi_{\mu\nu}\phi_{\rho\sigma} \,, \label{DHOST_action}
\end{align}
where
\begin{align}
C^{\mu\nu,\rho\sigma}&=\alpha_1 g^{\rho(\mu}g^{\nu)\sigma}
+\alpha_2 g^{\mu\nu}g^{\rho\sigma}
\nn
&+\frac{1}{2}\alpha_3(\phi^{\mu}\phi^{\nu}g^{\rho\sigma}+\phi^{\rho}\phi^{\sigma}g^{\mu\nu})
\nn 
&+\frac{1}{2}\alpha_4(\phi^{\rho}\phi^{(\mu}g^{\nu)\sigma}+\phi^{\sigma}\phi^{(\mu}g^{\nu)\rho} )+\alpha_5 \phi^{\mu}\phi^{\nu}\phi^{\rho}\phi^{\sigma}\,.
\end{align}
The explicit form of $\kappa$ is written in Appendix \ref{app_sol}.
The coefficients are given by
\begin{align}
f&=f_1-\frac{1}{2}f_2X\,, \label{coe_f} \\
P&=F_2+\frac{3X(f_{1\phi}-F_3X)^2}{2f_1-f_2X+2F_4X^2}
\,, \\
Q_1&=-2f_{\phi}+\frac{4f_1 (f_{1\phi}-F_3 X)}{2f_1-f_2X+2F_4X^2}
\,, \\
Q_2&=\frac{2f_{\phi}}X-\frac{4(f_1-3f_{1X}) (f_{1\phi}-F_3 X)}{X(2f_1-f_2X+2F_4X^2)}
 \,, \\
\alpha_1&=-\alpha_2=-\frac{f_2}{2}-\frac{f_1(f_2-2F_4X)}{2f_1-f_2X+2F_4X^2}
\,, \\
\alpha_3&=2f_{2X}+\frac{4f_1F_4+(4f_{1X}-f_2)(f_2-2F_4 X)}{2f_1-f_2X+2F_4X^2}
\,, \\
\alpha_4&=-2f_{2X}+2f_1^{-1}f_{1X}(3f_{1X}-f_2)
\nn
&+f_1^{-2}f_{1X}X(f_{1X}f_2-4f_1 f_{2X})
\nn
&+\frac{f_2^2-4f_1F_4-2f_2F_4X}{2f_1-f_2X+2F_4X^2}
\,, \\
\alpha_5&=- f_1^{-2} f_{1X}(f_{1X} f_2-4f_1 f_{2X})
\nn
&+\frac{2f_{1X}\{ 4f_1 F_4+(3f_{1X}-f_2)(f_2-2F_4 X)\}}{f_1(2f_1-f_2X+2F_4X^2)} \,, \label{coe_5}
\end{align}
where $f_{1\phi}=\partial f_1/\partial \phi,f_{1X}=\partial f_1/\partial X$ and so on. One can observe that \eqref{coe_f}-\eqref{coe_5} satisfy the degeneracy conditions. The resultant action is class ${}^2$N-I/Ia of quadratic DHOST. This class depends on five arbitrary functions which is indeed the same number of the arbitrary functions of \eqref{generalized_g}.

While the totally antisymmetric part $\kappa^{[\alpha\beta\gamma]}$ is zero, the antisymmetric trace $\kappa^{[\alpha\beta]}{}_{\beta}$ is non-zero and then \eqref{generalized_g} yields non-minimal couplings to the Dirac field \eqref{Dirac2}. In this sense, \eqref{generalized_g} is equivalent to class ${}^2$N-I/Ia of DHOST only if we do not consider the Dirac field and \eqref{generalized_g} is potentially a theory beyond DHOST due to a coupling to the Dirac field given by \eqref{int}. The trace of the antisymmetric part of $\kappa$ is
\begin{align}
&\kappa^{[\alpha\beta]}{}_{\beta}
\nn
=&
-\frac{3(f_{1\phi}-F_3 X)}{2f_1-f_2 X +2 F_4X^2} \phi^{\alpha}
-\frac{f_2-2F_4 X }{2f_1-f_2 X +2 F_4X^2} \phi^{\alpha} \phi^{\beta}_{\beta}
\nn
&-\frac{1}{2f_1^2}(f_{1X}f_2 X+6f_1 f_{1X}-f_1 f_2-2 f_1 f_{2X} X ) \phi^{\alpha \beta}\phi_{\beta} 
\nn
&+\frac{1}{2f_1^2}\Bigl[ f_{1X}f_2-2f_1 f_{2X} 
\nn
&\quad
- \frac{f_1\{ 4f_1 F_4+(6f_{1X}-f_2)(f_2-2F_4 X) \} }{2f_1-f_2 X+2 F_4 X^2} \Bigl] \phi^{\alpha} \phi^{\beta} \phi^{\gamma} \phi_{\beta\gamma}
\,,
\end{align}
which indicates that generic scalar-tensor theories may predict the non-minimal coupling between $\phi$ and fermions in the metric-affine formalism as well.

Note that the action \eqref{generalized_g} is not the most general projective invariant action up to quadratic in the connection. We can find more general Lagrangian by only assuming the projective invariance; however, such generalized theories are suffered from the Ostrogradsky instability which is discussed in Appendix \ref{appendix}. A typical example is a term $(\mathcal{L}^{\rm gal \Gamma}_3)^2$ which is certainly projective invariant but it leads to the Ostrogradky ghost. Therefore, we cannot unfortunately conclude that the structure of DHOST is protected by the projective invariance.

\subsection{Specific models}

We shall discuss some specific models. The Galileon field has been already discussed in the previous section. Here, we consider non-minimal couplings to the curvature. 
One of the simplest models of a non-minimal scalar field is
\begin{align}
\mathcal{L}=\frac{M_{\rm pl}^2-\xi \phi^2}{2}g^{\mu\nu}\RA{}_{\mu\nu}-\frac{1}{2}(\partial \phi)^2  -V(\phi)\,. \label{phi^2R}
\end{align}
In particular, in the case of the metric formalism, the non-minimal coupling $\xi \phi^2 R$ with $\xi=1/6$ is known as the conformal coupling. If the coupling $\xi \phi^2 R$ exists in the metric-affine formalisms, integrating out $\kappa$, the Lagrangian \eqref{phi^2R} becomes 
\begin{align}
\mathcal{L}=\frac{M_{\rm pl}^2-\xi \phi^2}{2}R(g)-\frac{M_{\rm pl}^2-\xi(1+6\xi )\phi^2}{2(M_{\rm pl}^2-\xi \phi^2)}(\partial \phi)^2  -V(\phi)\,,
\end{align}
where $\xi=1/6$ is no longer the conformal coupling due to the non-canonical kinetic term. A similar action was first considered in~\cite{Bauer:2008zj} for a torsionless case, and our results agree with their result when transformed into the Einstein frame.

Another example is the non-minimal coupling to the Einstein tensor,
\begin{align}
\mathcal{L}=\frac{M_{\rm pl}^2}{2}g^{\mu\nu}\RA{}_{\mu\nu}-\frac{1}{2}\left(g^{\mu\nu}-\frac{\GA{}^{\mu\nu}}{M^2} \right) \partial_{\mu} \phi \partial_{\nu} \phi -V(\phi)\,.
\label{G_coupling}
\end{align}
In the metric formalism, the Einstein tensor coupling $G^{\mu\nu}\partial_{\mu}\phi \partial_{\nu}\phi$ is in the class of the Horndeski theory. On the other hand, we obtain
\begin{align}
\mathcal{L}&=\frac{M_{\rm pl}^2}{2}R(g)-\frac{1}{2}\left(g^{\mu\nu}-\frac{G^{\mu\nu}(g)}{M^2} \right) \partial_{\mu} \phi \partial_{\nu} \phi -V(\phi)
\nn
&-\frac{1}{4M^4M_{\rm pl}^2(2-X/M^2M_{\rm pl}^2)} \mathcal{L}_4^{{\rm gal}g}\,,
\end{align}
from \eqref{G_coupling} after integrating out $\kappa$,
which is in the class of the GLPV theory due to the quartic Galileon term $\mathcal{L}_4^{{\rm gal}g}$. 

A theory in the class of the DHOST theory is obtained by considering the kinetic coupling to the curvature. For instance, let us assume that the scalar field appears only through $X=(\partial \phi)^2$ in the action. The most general action of the form $\mathcal{L}=\mathcal{L}(g,\Gamma,X)$ up to linear in the curvature is
\begin{align}
\mathcal{L}=f(X)g^{\mu\nu}\RA{}_{\mu\nu}+P(X)\,, \label{P(X)}
\end{align}
whose equivalent action in the metric formalism is
\begin{align}
\mathcal{L}=fR(g)+P+\frac{6f_{X}^2}{f}\phi^{\alpha}\phi^{\beta}\phi_{\alpha\gamma}\phi_{\beta}^{\gamma}\,,
\label{P(X)_metric}
\end{align}
which is in the class of the DHOST theory. 

We note that \eqref{P(X)_metric} automatically has the structure $\alpha_1=\alpha_2=0$. The modified gravity theories to explain the present cosmic accelerating expansion are strongly constrained by the speed of gravitational wave~\cite{TheLIGOScientific:2017qsa,Monitor:2017mdv}. In order that the speed of the gravitational wave exactly coincides with the speed of light, the functions in DHOST should be $f=f(\phi)$ and $\alpha_i=0$ or should be fine-tuned to $\alpha_1=\alpha_2=0$ with $f=f(\phi,X)$ ~\cite{Creminelli:2017sry,Sakstein:2017xjx,Ezquiaga:2017ekz} (see also \cite{Crisostomi:2017lbg,Langlois:2017dyl,Dima:2017pwp}). However, \eqref{P(X)_metric} does not require the fine-tuning even with a non-minimal coupling $f(X)R$ since \eqref{P(X)_metric} is obtained from the simple action \eqref{P(X)} where the ``counterterm'' $\phi^{\alpha}\phi^{\beta}\phi_{\alpha\gamma}\phi_{\beta}^{\gamma}$ to eliminate the Ostrogradsky ghost, which does not change the speed of gravitational waves, is automatically obtained by integrating out the distortion tensor in the action \eqref{P(X)}.

\section{Higher orders of connection}
\label{sec_higher_orders}
So far, we have considered theories up to quadratic in the connection in order to explicitly solve the equation of motion of the connection. When the Lagrangian $\mathcal{L}(g,\Gamma,\phi,\nablaA \phi, \nablaA \nablaA \phi)$ contains terms cubic or higher in the connection, a solution of the connection may be given by
\begin{align}
\kappa^{\mu}{}_{\alpha\beta} = \sum^{\infty}_{i,j,k}k^k_{i,j}(\phi,X) [(\nabla \phi)^i (\nabla \nabla \phi)^j]^{\mu}{}_{\alpha\beta}\,,
\end{align}
where the label $k$ classifies possible contractions of $(\nabla \phi)^i (\nabla \nabla \phi)^j$ with the free indices $\mu,\alpha,\beta$ for the same $i$ and $j$. Up to $j=1$, we obtain
\begin{align}
\kappa^{\mu}{}_{\alpha\beta} &=
k_{1,0}^{1}g_{\alpha\beta} \phi^{\mu} +k_{1,0}^2\delta^{\mu}_{\alpha}\phi_{\beta}+k_{1,0}^3 \delta^{\mu}_{\beta}\phi_{\alpha}+k_{3,0}^1\phi^{\mu}\phi_{\alpha}\phi_{\beta}
\nn
&+k_{1,1}^1 g_{\alpha\beta}\phi^{\mu}\phi^{\gamma}_{\gamma}+k_{1,1}^2 g_{\alpha\beta}\phi_{\gamma}\phi^{\mu\gamma}+k^3_{1,1}\delta^{\mu}_{\alpha}\phi_{\beta}\phi^{\gamma}_{\gamma}
\nn
&+k^4_{1,1}\delta^{\mu}_{\beta}\phi_{\alpha}\phi^{\gamma}_{\gamma}+k^5_{1,1}\delta^{\mu}_{\alpha}\phi^{\gamma}\phi_{\beta\gamma}+k^6_{1,1}\delta^{\mu}_{\beta}\phi^{\gamma}\phi_{\alpha\gamma}
\nn
&+k^7_{1,1}\phi^{\mu}\phi_{\alpha\beta}+k^8_{1,1}\phi_{\alpha}\phi^{\mu}_{\beta}+k^9_{1,1}\phi_{\beta}\phi^{\mu}_{\alpha}
\nn
&+k^1_{3,1}g_{\alpha\beta}\phi^{\mu}\phi^{\gamma}\phi^{\delta}\phi_{\gamma\delta}+k^2_{3,1}\delta^{\mu}_{\alpha}\phi_{\beta}\phi^{\gamma}\phi^{\delta}\phi_{\gamma\delta}
\nn
&+k^3_{3,1}\delta^{\mu}_{\beta}\phi_{\alpha}\phi^{\gamma}\phi^{\delta}\phi_{\gamma\delta}
+k^4_{3,1}\phi^{\mu}\phi_{\alpha}\phi_{\beta}\phi^{\gamma}_{\gamma}
\nn
&+k^5_{3,1}\phi^{\mu}\phi_{\alpha}\phi^{\gamma}\phi_{\beta\gamma}+k^6_{3,1}\phi^{\mu}\phi_{\beta}\phi^{\gamma}\phi_{\alpha\gamma}
\nn
&+k^7_{3,1}\phi_{\alpha}\phi_{\beta}\phi_{\gamma}\phi^{\mu\gamma}+k^1_{5,1}\phi^{\mu}\phi_{\alpha}\phi_{\beta}\phi^{\gamma}\phi^{\delta}\phi_{\gamma\delta}
+\mathcal{O}(\phi_{\mu\nu}^2)
\,. \label{kappa}
\end{align}
Note that the expression \eqref{kappa} contains the projective mode $k^2_{1,0},k^3_{1,1},k^5_{1,1},k^2_{3,1}$ and they can be removed by the projective transformation. Then, \eqref{kappa} has 17 independent terms up to linear in the second derivative of $\phi$.

Since the second derivative of the scalar field is given by
\begin{align}
\nablaA{}_{\mu}\nablaA{}_{\nu}\phi=\phi_{\mu\nu} -\kappa^{\alpha}{}_{\nu\mu} \phi_{\alpha}
\,,
\end{align}
the second derivative $\nablaA{}_{\mu}\nablaA{}_{\nu}\phi$ is still linear in $\phi_{\alpha\beta}$ if the equation of motion of $\kappa$ admits a solution
\begin{align}
k^k_{i,j}=0 \quad {\rm for} \quad  j \geq 2\,. \label{linear_kappa}
\end{align}
We thus consider whether the quintic Galileon \eqref{galA} can admit the solution \eqref{linear_kappa} by tuning the remaining 17 coefficients $k^k_{i,0}$ and $k^k_{i,1}$. After a straightforward calculation, we find that \eqref{galA} does not admit the solution \eqref{linear_kappa} if $c_5\neq 0$. This result implies that, when formally integrating out $\kappa$, the quintic Galileon \eqref{galA} must generate terms more than cubic in $\phi_{\mu\nu}$ and then the resultant theory cannot be in the cubic DHOST theory except for the case when there are miracle cancellations in higher orders of $\phi_{\mu\nu}$. Therefore, it would be interesting to investigate whether the quintic Galileon is still ghost-free even in the metric-affine formalism which is nonetheless beyond the scope of the present paper.

\section{Summary and discussions}
\label{sec_summary}
In the present paper, we have reformulated scalar-tensor theories in the metric formalism to those in the metric-affine formalism and clarified the relation between them. We assume that the Lagrangian is projective invariant and that the scalar field does not directly couple with the non-metricity tensor. Then, we find that the covariant Galileon terms in the metric-affine formalism are uniquely specified at least up to the quartic order although those in the metric formalism are not. The covariant Galileon in the metric-affine formalism does not coincide with either of those in the metric formalism where the deviation becomes relevant at the scales beyond $\Lambda_2$. 
%Another difference between the metric formalism and the metric-affine formalism is that the Galileon must non-minimally couple with fermionic fields in the metric-affine formalism. 
We then discuss a straightforward generalization of the Galileon and obtain an equivalent action to class ${}^2$N-I/Ia of DHOST. The equivalent action \eqref{generalized_g} makes the structure of DHOST clear because it is just linear in the generalized Galileon terms and the non-minimal couplings to the Ricci curvature and the Einstein tensor. We should, however, emphasize that the equivalence between \eqref{generalized_g} and class ${}^2$N-I/Ia of DHOST holds only if we ignore the fermionic fields. 

An important difference between theories in the metric formalism and in the metric-affine formalism is the coupling to the fermions. When we a priori assume the torsionless condition and the metric compatibility condition, i.e., $\kappa=0$, scalar-tensor theories should predict the universal coupling to bosons and fermions. On the other hand, as clarified in Section \ref{sec_metric_affine}, the standard bosonic fields do not couple with the distortion tensor $\kappa$ while the fermionic fields couple with $\kappa$ in the metric-affine formalism where the interaction depends on the definitions of the Dirac field Lagrangian. Although GR leads to the solution $\kappa=0$, in general, scalar-tensor theories yield $\kappa^{[\alpha\beta]}{}_{\beta} \neq 0$ and lead to the non-minimal coupling only with the fermions in the case of \eqref{Dirac2}.  Hence, it would be interesting to study phenomenological signatures of this non-minimal coupling.

The metric-affine formalism of gravity is sometimes called the first order formalism of gravity because the curvature is given by the first order derivative of the connection. The Ostrogradsky ghost-freeness of \eqref{generalized_g} could be understood by the fact that \eqref{generalized_g} does not contain second order derivatives of the fields except the Galileon combinations. To obtain the effective description of \eqref{generalized_g} in the metric formalism, one should decompose the general connection $\Gamma$ to the Levi-Civita connection and the distortion tensor $\kappa$, and integrate out $\kappa$; then, the Lagrangian \eqref{generalized_g} becomes to contain the non-Galileon combinations of the second order derivatives. However, the resultant Lagrangian is Ostrogradsky ghost-free which can be seen to satisfy the degeneracy condition. Needless to say, the non-existence of the ghost is not obvious even if a theory only contains first order derivatives because one may reduce the number of the derivatives by introducing an auxiliary field with a Lagrangian multiplier. Nonetheless, the metric-affine formalism could give a new understanding of the ghost-freeness of DHOST: the non-Galileon combinations of the second derivatives of DHOST are rewritten by the first order form in the metric-affine formalism.

Although we have discussed theories up to quadratic order in the connection in order to solve $\kappa$ explicitly, one may discuss higher-order theories with respect to the connection. The simplest theory would be the quintic Galileon and its generalization. Other possible extensions would be the Fab Four type Lagrangian ~\cite{Charmousis:2011bf}. Since the first two terms of \eqref{generalized_g} are two of Fab Four, we may expect that the last two of Fab Four
\begin{align}
f_3(\phi,X) \GA{}^{\mu\alpha\nu\beta}\nablaA{}_{\mu}\phi \nablaA{}_{\nu}\phi \nablaA{}_{\alpha}\nablaA{}_{\beta}\phi \,, ~
f_4(\phi,X) \GA{}^{\mu\nu \alpha\beta} \RA{}_{\mu\nu\alpha\beta}
\end{align}
are ghost-free as well where $\GA{}^{\mu\alpha\nu\beta}$ is the double dual of the Riemann curvature defined by \eqref{ddR}. It would be worthwhile to discuss whether the quintic Galileon and Fab Four are still ghost-free in the metric-affine formalism, which is left for future works.

%%%%%%%%%%%%%%%%%%%%%%%%%%%%%%%%%%%%%%%%%%%%%%%%%%%%%%%%%%%%%%%
%%%%%%%%%%%%%%%%%%%%%%%%%%%%%%%%%%%%%%%%%%%%%%%%%%%%%%%%%%%%%%%
\section*{Acknowledgments}
%%%%%%%%%%%%%%%%%%%%%%%%%%%%%%%%%%%%%%%%%%%%%%%%%%%%%%%%%%%%%%%
%%%%%%%%%%%%%%%%%%%%%%%%%%%%%%%%%%%%%%%%%%%%%%%%%%%%%%%%%%%%%%%
We would like to thank Kei-ichi Maeda and Yusuke Yamada for useful discussions and comments.
The work of K.A. was supported in part by Grants-in-Aid from the Scientific Research Fund of the Japan Society for the Promotion of Science  (No.~15J05540) and by a Waseda University Grant for Special Research Projects (No.~2018S-128).

%%%%%%%%%%%%%%%%%%%%%%%%%%%%%%%%%%%%%%%%%%%%%%%%%%%
%%%%%%%%%%%%%%%%%%%%%%%%%%%%%%%%%%%%%%%%%%%%%%%%%%%
%%%%%%%%%%%%%%%%%%%%%%%%%%%%%%%%%%%%%%%%%%%%%%%%%%%
%%%%%%%%%%%%%%%%%%%%%%%%%%%%%%%%%%%%%%%%%%%%%%%%%%%
%%%%%%%%%%%%%%%%%%%%%%%%%%%%%%%%%%%%%%%%%%%%%%%%%%%
%%%%%%%%%%%%%%%%%%%%%%%%%%%%%%%%%%%%%%%%%%%%%%%%%%%
%%%%%%%%%%%%%%%%%%%%%%%%%%%%%%%%%%%%%%%%%%%%%%%%%%%

\appendix

\section{Projective invariant scalar-tensor theories}
\label{appendix}
We consider scalar-tensor theories in the metric-affine formalism whose action is constructed by the metric tensor $g_{\mu\nu}$, the curvature tensor $\RA{}^{\mu}{}_{\nu\alpha\beta}$, the scalar field $\phi$, and its covariant derivatives. For simplicity, we only consider the action up to quadratic in the connection. The projective invariance leads to
\begin{align}
\mathcal{L}&=f g^{\mu\nu} \RA{}_{\mu\nu} + g_1 g^{\mu\alpha}g^{\nu\beta} \RA{}_{\mu\nu} \partial_{\alpha} \phi \partial_{\beta} \phi 
\nn&
+ g_2 g^{\alpha\beta} g^{\mu\nu} \RA{}^{\rho}{}_{\alpha\mu\beta} \partial_{\rho} \phi \partial_{\nu}\phi
+F_2 +F_3 \mathcal{L}_3^{\rm gal \Gamma}
+F_4 \mathcal{L}_4^{\rm gal \Gamma}
\nn
&+C_1 \epsilon^{\mu\nu\rho\sigma}\epsilon^{\mu'\nu'\rho'}{}_{\sigma}\partial_{\mu}\phi \partial_{\mu'} \phi \nablaA{}_{\nu}\nablaA{}_{\nu'}\phi \nablaA{}_{[\rho}\nablaA{}_{\rho']}\phi
+C_2 ( \mathcal{L}_3^{\rm gal \Gamma})^2
\nn
&+C_3 (g^{\mu\beta}g^{\nu\delta}g^{\alpha\gamma}-g^{\mu\nu}g^{\alpha\gamma}g^{\beta\delta})\partial_{\mu}\phi \partial_{\nu} \phi \nablaA{}_{\alpha}\nablaA{}_{\beta}\phi \nablaA{}_{\gamma}\nablaA{}_{\delta} \phi\,,
\label{affine_action}
\end{align}
where $f,g_1,g_2,F_2,F_2,F_4,C_1,C_2,C_3$ are arbitrary functions of $\phi$ and $X:=g^{\mu\nu}\partial_{\mu}\phi \partial_{\nu} \phi$. The solution of $\kappa$ is give by the form \eqref{kappa} with \eqref{linear_kappa}.
We note that $k^k_{i,j}$ are expressed in terms of $f,g_1,g_2,F_2,F_3,F_4,C_2,C_3$ but they have no dependence on $C_1$. Furthermore, we also find that the relation between $(P,Q_1,Q_2,\alpha_i)$ of the action \eqref{DHOST_action} and $(f,g_1,g_2,F_2,F_2,F_4,C_1,C_2,C_3)$ where $P,Q_1,Q_2,\alpha_i$ are independent from $C_1$. Therefore, the following Galileon terms
\begin{align}
\epsilon^{\alpha\beta\gamma\delta} \epsilon^{\alpha'\beta'\gamma'}{}_{\delta} \nablaA{}_{\alpha} \phi \nablaA{}_{\alpha'} \phi \nablaA{}_{\beta} \nablaA{}_{\beta'}\phi \nablaA{}_{\gamma} \nablaA{}_{\gamma'} \phi
\end{align}
and
\begin{align}
\epsilon^{\alpha\beta\gamma\delta} \epsilon^{\alpha'\beta'\gamma'}{}_{\delta} \nablaA{}_{\alpha} \phi \nablaA{}_{\alpha'} \phi \nablaA{}_{\beta} \nablaA{}_{\beta'}\phi \nablaA{}_{\gamma'} \nablaA{}_{\gamma} \phi
\end{align}
give the same result although $\nablaA{}_{\gamma} \nablaA{}_{\gamma'} \phi$ is not symmetric for the indices $\gamma$ and $\gamma'$.

The Ostrogradsky ghost-free conditions of the action \eqref{DHOST_action} are given by
\begin{align}
D_0=0\,,~D_1=0\,,~D_2=0\,,
\end{align}
where
\begin{align}
D_0&:=-4(\alpha_1+\alpha_2)
\nn 
& \qquad \times [X f(2\alpha_1+X\alpha_4+4 f_{X})-2f^2-8X^2 f_{X}^2]
\,, \\
D_1&:=4[X^2 \alpha_1 (\alpha_1+3\alpha_2)-2f^2-4X f \alpha_2]\alpha_4 
\nn
&~ +4 X^2 f(\alpha_1+\alpha_2)\alpha_5 +8 X \alpha_1^3 
\nn
&~-4 (f+4 X f_X -6X \alpha_2)\alpha_1^2 -16 (f+5X f_X) \alpha_1 \alpha_2
\nn
&~  +4 X (3f-4X f_X) \alpha_1 \alpha_3 -X^2 f \alpha_3^2 
\nn 
&~ +32 f_X (f+2 X f_X)\alpha_2 -16 f f_X \alpha_1 
\nn
&~ -8f(f-X f_X) \alpha_3 +48 f f_X^2
\,,
\\
D_2&:=4[2f^2+4X f \alpha_2 -X^2 \alpha_1(\alpha_1+3\alpha_2)]\alpha_5 +4\alpha_1^3 
\nn 
&~ +4 (2\alpha_2 -X \alpha_3-4f_X) \alpha_1^2+3 X^2 \alpha_1 \alpha_3^2 -4X f \alpha_3^2
\nn
&~+8(f+X f_X) \alpha_1 \alpha_3 -32 f_X \alpha_1 \alpha_2 +16 f_X^2 \alpha_1 
\nn 
&~+32 f_X^2 \alpha_2 -16 f f_X \alpha_3\,.
\end{align}
Following the classification \cite{Achour:2016rkg}, the case 
\begin{align}
\alpha_1+\alpha_2=0 \,,
\end{align} 
is called class I and the case 
\begin{align}
X f(2\alpha_1+X\alpha_4+4 f_{X})-2f^2-8X^2 f_{X}^2=0 \,,~ f\neq 0
\end{align}
is called class II, respectively. Class I (or class II) can be further classified into class Ia (or class IIa) if $f\neq X \alpha_1$ and class Ib (or class IIb) if $f=X \alpha_1$. A special class $f=0$ is called class III. As for the theory \eqref{affine_action}, $D_i$ are given by
\begin{align}
D_0&=-\frac{8X(C_3-4C_2 X)(f+g_2 X)^2 D^2}{E}\,, 
\\
D_1&=-\frac{8(C_3-8C_2 X)(f+g_2 X)^2 D^2}{E} \,,
\\
D_2 &= \frac{32 C_2 (f+g_2 X)^2 D^2}{E}\,,
\end{align}
where 
\begin{align}
D&:=2f g-4f_X g X +f g_X X
\,,
\\
E&:=[2f^2+2g_2^2 X^2 + f X (4g_2-C_3 X)] 
\nn
&~\times [f^2+(F_4-C_3) g_1 X^3 +f X \{ g_2 +(F_4-C_3)X\} ]
\nn
&~\times [ 2f^2+g_1 X^3(2F_4+C_3 -12 C_2 X)
\nn
& \quad \quad +f X (2g_2 +X \{ 2F_4+C_3-12 C_2 X \}]\,.
\end{align}
with $g:=(f+g_1 X)(f+g_2 X)$.
The Ostrogradsky ghost-free theories of \eqref{affine_action} are classified into four cases
\begin{align}
{\rm class~Ia}:&~C_2=C_3=0\,, \\
{\rm class~IIa}:&~D=0\,,~g\neq 0 \,, ~f\neq 0 \,,\\
{\rm class~Ib}\cap {\rm IIb}:&~g=0\\
{\rm class~III}:&~f=0\,,
\end{align}
which correspond to the classification of DHOST, respectively. Only class Ia is obtained by eliminating the non-Galileon terms of the derivative self-interactions of the scalar field while other classes are obtained by tuning the curvature couplings.

Since the phenomenologically viable class is only class Ia~\cite{Langlois:2017mxy}, we shall focus on only the case $C_2=C_3=0$. In this case, the functions are explicitly given by
\begin{align}
P&= F_2+\frac{3X (g_{\phi}-2F_3' X)^2}{8 (f g +F_4' X^2)}
\,, \\
Q_1 &= -2f_{\phi}+\frac{g (g_{\phi}-2F_3' X)}{fg + F_4' X^2}
\,, \\
Q_2 &= \frac{2 f_{\phi}}{X}-\frac{(g_{\phi}-2 F_3' X)(2g -3 g_X X)}{2 X (f g +F_4' X^2)}
\,, \\
\alpha_1&=-\alpha_2 =\frac{1}{X} \left[ f - \frac{g^2}{f g +F_4' X^2} \right]
\,, \\
\alpha_3&=\frac{2}{X^2}\left[ f -2 f_X X -\frac{g(g-g_X X)}{f g +F_4' X^2} \right]
\,, \\
\alpha_4 &=\frac{1}{2g^2 X^2} [ 8 f_X g X(g+g_X X) -f(2g +g_X X)^2]
\nn &
+\frac{2g^2}{fg X^2 +F_4' X^4}
\,, \\
\alpha_5&=\frac{g_X}{2X^2g^2 } (4f g - 8 f_X g X +f g_X X)
\nn
& - \frac{g_X (4g -3g_X X)}{2X^2(fg +F_4' X^2)} \,,
\end{align}
where 
\begin{align}
F_3':=F_3 (f+g_1 X) \,, ~
F_4':= F_4 (f+g_1 X)^2
\,.
\end{align}
While the action \eqref{affine_action} with $C_2=C_3=0$ contains six independent functions $f,g_1,g_2,F_2,F_3,F_4$, the equivalent action \eqref{DHOST_action} is specified by only five combinations $f,g,F_2,F_3',F_4'$.

We note that the double dual of the Riemann curvature,
\begin{align}
\GA{}^{\mu\nu\alpha\beta}&:= \frac{1}{4} \epsilon^{\mu\nu \rho\sigma  }\epsilon^{\alpha\beta \rho\sigma  } \RA{}_{\rho\sigma\rho'\sigma'}
\,, \label{ddR}
\end{align}
is projective invariant although the Riemann curvature itself is not. The Ricci scalar and the Einstein tensor are given by contractions of $\GA{}^{\mu\nu\alpha\beta}$:
\begin{align}
g^{\mu\nu} \RA{}_{\mu\nu}  = - \GA{}_{\mu\nu}{}^{\mu\nu} 
\,, \quad
\GA{}^{\mu\nu} = \GA{}_{\alpha}{}^{\mu\alpha\nu}
\,. 
\end{align}
and then they are definitely projective invariant. Let us assume that the non-minimal couplings between $\phi$ and the curvature are given by the couplings to the double dual of the Riemann curvature instead of the Riemann curvature. This assumption leads to $g_1=g_2$ and then the theory \eqref{affine_action} with $C_2=C_3=0$ is reduced to \eqref{generalized_g}. 

\section{Explicit solution of $\kappa$}
\label{app_sol}
The solution of $\kappa$ to the theory \eqref{generalized_g} is given by the form \eqref{kappa} with 
\begin{align}
k^1_{1,0}&=-\frac{f_{1\phi}-F_3 X}{2f_1-f_2 X +2 F_4 X^2} \,,
\nn
k^3_{1,0}&=-k^1_{1,0}=\frac{f_{1\phi}-F_3 X}{2f_1-f_2 X +2 F_4 X^2} \,,
\end{align}
\begin{align}
k^1_{3,0}&=-\frac{2f_1 F_3-f_2 F_3 X + 2 f_{1\phi} F_4 X}{f_1 (2f_1-f_2 X +2 F_4 X^2)} \,,
\end{align}
\begin{align}
k^1_{1,1}&=0 \,,
\nn
k^2_{1,1}&=-\frac{f_{1X}}{f_1} \,,
\nn
k^4_{1,1}&=0 \,,
\nn
k^6_{1,1}&=-k^2_{1,1}=\frac{f_{1X}}{f_1} \,, 
\nn
k^7_{1,1}&=-\frac{f_2-2F_4 X}{2f_1-f_2 X +2 F_4 X^2} \,,
\nn
k^8_{1,1}&=-k^7_{1,1}=\frac{f_2-2F_4 X}{2f_1-f_2 X +2 F_4 X^2} \,,
\nn
k^9_{1,1}&=-\frac{2F_4 X}{2f_1-f_2 X +2 F_4 X^2} \,,
\end{align}
\begin{align}
k^1_{3,1}&=-\frac{f_{1X}(f_2-2F_4 X)}{f_1 (2f_1-f_2 X +2 F_4 X^2)} \,,
\nn
k^3_{3,1}&=-k^1_{3,1}=\frac{f_{1X}(f_2-2F_4 X)}{f_1 (2f_1-f_2 X +2 F_4 X^2)} \,,
\nn
k^4_{3,1}&=-\frac{2F_4}{2f_1-f_2 X +2 F_4 X^2} \,,
\nn
k^5_{3,1}&=0 \,,
\nn
k^6_{3,1}&=\frac{1}{2f_1^2}\left[ f_{1X} f_2-2f_1 f_{2X}+ \frac{f_1 f_2 (f_2 -2 F_4 X)}{2f_1-f_2 X +2 F_4 X^2} \right] ,
\nn
k^7_{3,1}&=-\frac{1}{2f_1^2}\biggl[ f_{1X} f_2-2f_1 f_{2X}
\nn 
&\qquad \qquad
+\frac{f_1 (f_2^2-8 f_1 F_4 -2f_2 F_4 X))}{2f_1-f_2 X +2 F_4 X^2} \biggl] \,,
\end{align}
\begin{align}
k^1_{5,1}&=-\frac{4 f_{1X} F_4}{f_1 (2f_1-f_2 X +2 F_4 X^2)} \,,
\end{align}
and
\begin{align}
k^{k}_{i,j}=0\,~{\rm for}~j \geq 2 \,.
\end{align}

\newpage
\bibliography{ref}
\bibliographystyle{JHEP}

\end{document}